\documentclass[aip,floatfix,twocolumn,showpacs,preprintnumbers,amsmath,amssymb,superscriptaddress,longbibliography,reprint]{revtex4-1}
\usepackage{color}
\usepackage[usenames,dvipsnames,svgnames,table]{xcolor}
\usepackage[colorlinks=true,linkcolor=blue,urlcolor=blue,citecolor=blue]{hyperref}

\usepackage{mathtools}
\usepackage{graphicx}
\usepackage{dcolumn}
\usepackage{array}
\usepackage{lipsum}
\usepackage{bm}
\usepackage{subfigure}
\usepackage{amssymb}
\usepackage{multirow}
\usepackage{tabularx}
\usepackage{amsmath}
\usepackage{braket}

\newcommand{\br}{\bm{r}}
\newcommand{\bq}{\bm{q}}

\renewcommand{\vec}[1]{\mathbf{#1}}

\begin{document}

\title{Assessing the accuracy  of hybrid exchange-correlation functionals for the density response of warm dense electrons}

\author{Zhandos A. Moldabekov}
\email{z.moldabekov@hzdr.de}
\affiliation{Center for Advanced Systems Understanding (CASUS), Helmholtz-Zentrum Dresden-Rossendorf (HZDR), D-02826 G\"orlitz, Germany}

\author{Mani Lokamani}
\affiliation{Information Services and Computing, Helmholtz-Zentrum Dresden-Rossendorf (HZDR), D-01328 Dresden, Germany}

\author{Jan Vorberger}
\affiliation{Insitute of Radiation Physics, Helmholtz-Zentrum Dresden-Rossendorf (HZDR), D-01328 Dresden, Germany}

\author{Attila Cangi}
\affiliation{Center for Advanced Systems Understanding (CASUS), Helmholtz-Zentrum Dresden-Rossendorf (HZDR), D-02826 G\"orlitz, Germany}

\author{Tobias Dornheim}
\affiliation{Center for Advanced Systems Understanding (CASUS), Helmholtz-Zentrum Dresden-Rossendorf (HZDR), D-02826 G\"orlitz, Germany}

\begin{abstract}

We assess the accuracy of common hybrid exchange-correlation (XC) functionals (PBE0, PBE0-1/3, HSE06, HSE03, and B3LYP) within Kohn–Sham density functional theory (KS-DFT) for the harmonically perturbed electron gas at parameters relevant for the challenging conditions of warm dense matter. Generated by laser-induced compression and heating in the laboratory, warm dense matter is a state of matter that also occurs in white dwarfs and planetary interiors. We consider both weak and strong degrees of density inhomogeneity induced by the external field at various wavenumbers. We perform an error analysis by comparing to exact quantum Monte-Carlo results.
In the case of a weak perturbation, we report the static linear density response function and the static XC kernel at a metallic density for both the degenerate ground-state limit and for partial degeneracy at the electronic Fermi temperature. Overall, we observe an improvement in the density response for partial degeneracy when the PBE0, PBE0-1/3, HSE06, and HSE03 functionals are used compared to the previously reported results for the PBE, PBEsol, LDA, AM05, and SCAN functionals; B3LYP, on the other hand, does not perform well for the considered system. 
Together with the reduction of self-interaction errors, this seems to be the rationale behind the relative success of the HSE03 functional for the description of the experimental data on aluminum and liquid ammonia at WDM conditions.
\end{abstract}

\maketitle

\section{Introduction}

Kohn-Sham density functional theory (KS-DFT) is the most popular first-principles electronic structure method both at ambient and extreme conditions.
This includes simulations of warm dense matter \cite{wdm_book, new_POP, dornheim_physrep18_0,fortov_review}, where KS-DFT is a widely used ~\cite{dynamic2,Mo_PRL_2018,Ramakrishna_PRB_2021} for the interpretation of X-Ray Thomson scattering experiments~\cite{siegfried_review,kraus_xrts,Dornheim_T_2022}.  More specifically, WDM is a state of matter characterized by high temperatures and densities and is generated 
using different techniques~\cite{falk_wdm}
at large research facilities like the European X-ray Free-Electron Laser (XFEL) \cite{Tschentscher_2017}, the National Ignition Facility (NIF) \cite{BH16, PhysRevLett.104.235003}, the Linac coherent light source at SLAC, \cite{Ofori_Okai_2018},  and the Sandia Z-machine \cite{matzen_pulsed-power-driven_2005}.
Both experimental and theoretical investigations of WDM are of high current interest due to its relevance for astrophysics, material science, and inertial confinement fusion \cite{PhysRevLett.104.235003,hu_ICF}.
Examples of astrophysical applications include white dwarfs \cite{Kritcher2020} and planetary interiors \cite{Kraus2017, PhysRevB.101.054301, SEJ14}.
In addition, WDM is relevant for the discovery of novel materials at extreme conditions  \cite{Kraus2017, Lazicki2021, Luetgert2021,PhysRevResearch.2.033366} and for hot-electron chemistry \cite{Brongersma2015}.

The rigorous understanding of the physics and chemistry of the WDM state is still in a stage of active development since the most widely used electronic structure calculations and diagnostic tools were developed for ambient conditions and require a consistent extension to high temperatures.
For example, in the case of the KS-DFT, the key ingredient for an accurate simulation is the exchange-correlation (XC) functional.
At ambient conditions, the quality of different XC functionals is tested by comparing them to the experimental data, e.g., to band gaps \cite{Pedro}, lattice constants \cite{Paier}, and atomization energies \cite{Becke}.

In WDM simulations it is a common practice to use ground-state XC functionals with thermal excitations included by the smearing of band occupation numbers according to the Fermi-Dirac distribution \cite{mermin_65, giuliani2005quantum}; this is often referred to as \emph{zero-temperature approximation} in the literature.
Although there has been a certain progress in the development of finite-temperature XC functionals \cite{PhysRevLett.107.163001, groth_prl, Karasiev_PRL_2018, PhysRevB.86.115101, PhysRevB.101.245141}, testing XC functionals at WDM conditions is challenging. First, there are significant uncertainties in the experimentally realized temperature and density~\cite{kraus_xrts, Dornheim_T_2022, Dornheim2022Physical}. Second, besides very recent path integral Monte Carlo (PIMC) simulation results for hydrogen at certain WDM parameters \cite{Bohme_PRL_2022,Bohme_PRE_2022}, exact simulation data for complex materials like compressed solids at WDM conditions \cite{Denoeud, Booth2015, PhysRevResearch.2.033366} that would enable reliable benchmarks of XC functionals are sparse.

This situation is highly unsatisfactory because KS-DFT results are often used for the interpretation of experimental observations \cite{PhysRevResearch.2.033366,Mo_PRL_2018} and as an input to other methods and models including machine-learning models \cite{PhysRevMaterials.6.040301}, XC kernels for linear-response time-dependent density functional theory calculations \cite{dft_kernel}, molecular dynamics with screened effective potentials~\cite{ceperley_potential, zhandos2, PhysRevE.98.023207, PhysRevLett.109.225001,  doi:10.1063/5.0097768, https://doi.org/10.1002/ctpp.202000176, https://doi.org/10.1002/ctpp.201700109}, and quantum hydrodynamics~\cite{Murillo,zhandos_pop18,Moldabekov_SciPost_2022}.

Clearly, an extensive benchmark and analysis of the accuracy of KS-DFT calculations based on various XC functionals at high temperatures is needed.
Recently, Moldabekov \textit{et al.} \cite{Moldabekov_JCP_2021, dft_kernel, Moldabekov_PRB_2022} have reported the analysis of the LDA \cite{LDA_PW}, PBE \cite{PBE}, PBEsol \cite{PBEsol}, AM05 \cite{PhysRevB.72.085108},  and SCAN \cite{SCAN} XC functionals at temperatures and densities relevant for WDM by considering an inhomogeneous electron gas in an external harmonic perturbation and comparing DFT results with exact PIMC reference data~\cite{dornheim_ML,PhysRevLett.125.085001,Dornheim_PRR_2021}.
This has provided important insights to the performance of these functionals at WDM conditions. For example, in contrast to the ground state calculations, it was found that at high temperatures SCAN performs worse than LDA, PBE, and PBEsol.
Therefore, it is not expected that the conclusions about the accuracy of XC functionals in the ground state~\cite{Clay_PRB_2014,Clay_PRB_2016} remain fully valid at WDM conditions.
Furthermore, for the examples of the warm dense electron gas and hydrogen, it has been pointed out that even finite-temperature versions of the LDA and GGA functionals---created to extend the corresponding ground versions to high temperatures---do not always lead to an improved results at WDM conditions \cite{dft_kernel}.

On the other hand, a generalized KS-DFT \cite{RevModPhys.80.3} with orbital-dependent XC functionals has emerged as one of the most promising tools in density-functional theory.
The most often used orbital-dependent XC functionals \cite{RevModPhys.80.3, PhysRevX.10.021040} belong to the class of the hybrid functionals, which entail a certain weighted combination of the exact Hartree-Fock exchange contribution and of a standard XC functional like the LDA or PBE \cite{Becke, doi:10.1063/1.472933}.
These orbital-dependent XC functionals naturally contain information about non-locality effects and reduce the self-interaction error \cite{doi:10.1021/acs.jctc.0c00585}.
Moreover, they provide a superior description of various material properties like lattice constants \cite{Paier} and atomization energies  \cite{Becke}.

Hybrid XC functionals---within generalized KS-DFT at finite temperature \cite{PhysRevB.81.155119, PhysRevLett.107.163001, PhysRevB.101.245141}---can potentially overcome some problems of finite-temperature XC functionals since, in addition to the reduction of the self-interaction error, they contain extra information about thermal effects through orbitals in the exact exchange part.
Therefore, in this work we extend the previous benchmark and analysis of the LDA, GGA, and meta-GGA functionals for the harmonically perturbed electron gas at WDM conditions \cite{Moldabekov_JCP_2021, Moldabekov_PRB_2022} to include a number of popular hybrid XC functionals. In our assessment, we consider the HSE06 \cite{Heyd2006, Krukau2006}, HSE03 \cite{Heyd2003}, PBE0 \cite{Adamo1999}, PBE0-1/3 \cite{Cortona2012}, and B3LYP \cite{Stephens1994} hybrid functionals.

First, we consider the weakly perturbed electron gas and present results for the static hybrid XC kernel $K_\textnormal{xc}(\mathbf{q})$ of the uniform electron gas (UEG) at both ambient and WDM conditions using the recent framework by Moldabekov \textit{et al.} \cite{dft_kernel}. In particular, $K_\textnormal{xc}(\mathbf{q})$ constitutes a wave-vector resolved measure for electronic XC effects and, therefore, is uniquely suited to benchmark the performance of XC functionals. In addition, the kernel constitutes key input for a gamut of applications such as quantum hydrodynamics~\cite{zhandos_pop18,Murillo} or the interpretation of XRTS experiments~\cite{Fortmann_PRE_2010,Preston,Dornheim_PRL_2020_ESA,Zan_PRE_2021}.

Second, we consider a strongly inhomogeneous, warm dense electron gas.
The analysis is carried out by comparing with results for the PBE functional and exact PIMC reference data. PIMC allows us to unambiguously quantify the accuracy of the considered hybrid XC functionals, whereas the comparison with the PBE functional gives us insight into physics behind the observed behavior of the hybrid functionals at different conditions.


The paper is organized as follows: we begin with providing the theoretical background in Sec.~\ref{s:theory};
the details of our KS-DFT and PIMC calculations are given in Sec.~\ref{s:setup}; the results of the calculations are presented in Sec.~\ref{s:results}; the paper is concluded in Sec.~\ref{s:end} with a summary of the main findings and a brief outlook on future research.


\section{Theoretical backgrounds}\label{s:theory}

\subsection{Hybrid XC functionals for systems at finite temperature}\label{s:hybrids}

Hybrid XC functionals are based on the \emph{adiabatic connection} formula \cite{Becke}. At finite temperature, the thermal adiabatic connection formula for the exchange-correlation functional, in terms of the potential energy of exchange-correlation, $U_{\rm xc}^{\lambda}$ \cite{PhysRevB.13.4274}, is given by \cite{PhysRevLett.107.163001}
\begin{equation}
    F_{\rm xc}=\int_0^{1}\mathrm{d}\lambda\, U_{\rm xc}^{\lambda}\,,
\end{equation}
where $\lambda$ is the electron-electron coupling parameter (interaction strength). 
In the $\lambda=0$ limit, $U_{\rm xc}^{\lambda=0}$ reduces to the finite temperature Hartree–Fock exchange, with $U_{\rm xc}^{\lambda=0}=F_{X}^{\rm HF}$ given by \cite{B812838C, PhysRevB.101.245141}
\begin{equation}
F_{X}^{\rm HF}[\rho_{\sigma}]=-\frac{1}{2}\sum_{\sigma=\uparrow, \downarrow}\int \mathrm{d}\vec r_1 \int \mathrm{d}\vec r_2 \frac{\left|\rho_{\sigma}\left(\vec r_1,\vec r_2\right)\right|^{2}}{r_{\rm 12}},
\end{equation}
where $r_{\rm 12}=\left| \vec r_1-\vec r_2\right|$ and $\rho_{\sigma}\left(\vec r_1,\vec r_2\right)$ is the exchange density,
\begin{equation}
    \rho_{\sigma}\left(\vec r_1,\vec r_2\right)=\sum_i f_{\rm i\sigma}\phi_{\rm i\sigma}\left(\vec r_1\right)\phi_{\rm i\sigma}^{*}\left(\vec r_2\right),
\end{equation}
with $f_{\rm i\sigma}$ being the orbital occupation number defined by the Fermi–Dirac distribution.

Within the hybrid XC functionals construction scheme, the $\lambda=1$ limit $U_{\rm xc}^{\lambda=1}$ is approximated by a standard XC density functional, $E_{\rm xc}^{\rm DF}[n]$, such as LDA or GGA \cite{Becke, doi:10.1063/1.472933} (where DF is an abbreviation for density functional).
In their essence, hybrid XC functionals are based on the approximation of the $U_{\rm xc}^{\lambda}$ in the form of a certain (usually polynomial) interpolation between the limits $U_{\rm xc}^{\lambda=0}=F_{x}^{\rm HF}[\rho_{\sigma}]$ and $U_{\rm xc}^{\lambda=1}=E_{\rm xc}^{\rm DF}[n]$ \cite{Becke, doi:10.1063/1.472933, Adamo1999, Cortona2012, Stephens1994}. For example, employing \cite{doi:10.1063/1.472933}
\begin{equation}
    U_{\rm xc}^{\lambda}=E_{\rm xc}^{\rm DF}[n]+\left(F_{x}^{\rm HF}[\rho_{\sigma}]-E_{\rm x}^{\rm DF}[n]\right)\left(1-\lambda\right)^{n-1}
\end{equation}
with $n=4$ and $E_{\rm x}^{\rm DF}$ being the exchange part of $E_{\rm xc}^{\rm DF}$, and implementing the PBE as the DF, one arrives at a finite temperature version of the PBE0 hybrid XC functional \cite{Adamo1999}
\begin{equation}\label{eq:pbe0}
    F_{\rm xc}^{\rm PBE0}[\rho_{\sigma},n]=E_{\rm xc}^{\rm PBE}[n]+\frac{1}{4}\left(F_{x}^{\rm HF}[\rho_{\sigma}]- E_{\rm x}^{\rm PBE}[n]\right).
\end{equation}

In Eq. (\ref{eq:pbe0}), the implicit dependence on the electronic temperature is taken into account in $F_{x}^{\rm HF}[\rho_{\sigma}]$ due to the smearing of the occupation numbers.
Similarly, the usage of orbitals in the Fermi–Dirac distribution of the occupation numbers in the PBE0-1/3, HSE06, HSE03, and B3LYP allows for incorporating the implicit portion of thermal effects.  Clearly, this is not a fully consistent way of generating a finite-temperature hybrid XC functional. Nevertheless, e.g., the HSE03 has been successfully used for the description of the experimental X-ray spectra of aluminum \cite{PhysRevLett.118.225001} and the reflectivity and dc electrical conductivity of liquid ammonia \cite{PhysRevLett.126.025003} at high temperatures. Furthermore, it turns out that the straightforward use of the orbitals with smeared occupation numbers according to the Fermi–Dirac distribution in the PBE0, PBE0-1/3, HSE06, HSE03 provides a significant improvement in the description of the electronic density response at high temperatures $T\sim T_F$ compared to PBE as reported in Sec. \ref{s:results}.
Note that in the case of the HSE06 and HSE03, the Coulomb potential in $F_{x}^{\rm HF}[\rho_{\sigma}]$ is substituted by a screened potential $1/r_{\rm 12}\to {\rm erfc}(\Omega r_{\rm 12})/r_{\rm 12}$ \cite{B812838C} to avoid the computational difficulties due to the long-range character of the Coulomb potential. Consistently, the screened version for the exchange part of the PBE is also used. For example, $\Omega=0.11~{\rm Bohr^{-1}}$ is used for the HSE03 \cite{Krukau2006}.

\subsection{Density response functions and hybrid XC kernel}\label{s:t0_ueg}

We investigate the performance of various hybrid XC functionals at WDM conditions by analyzing the density response of the interacting electron gas governed by the following  Hamiltonian ~\cite{PhysRevLett.75.689, Dornheim_PRR_2021, SPB2016:exact}:
\begin{align}\label{eq:H}
\hat{H} = \hat{H}_{\rm UEG} +  \sum_{k=1}^N 2\, A \cos(\hat{\br}_k \cdot {\bq})\ ,
\end{align}
where $\hat{H}_{\rm UEG}$ is the Hamiltonian of the unperturbed UEG~\cite{loos,giuliani2005quantum,dornheim_physrep18_0}, and $N$ is the number of electrons. {Note that we restrict ourselves to the unpolarized case with an equal number of spin-up and spin-down electrons, $N^\uparrow=N^\downarrow=N/2$, throughout this work.}
Additionally to $\hat{H}_{\rm UEG}$, the Hamiltonian in Eq.~(\ref{eq:H}) has an extra harmonic perturbation defined by the amplitude $A$ and wavelength $\lambda=2\pi/|\vec q|$. 

It allows us to investigate both the weakly and strongly inhomogeneous electron gas by tuning the external perturbation.
As a measure of the density inhomogeneity, one can use the maximum value of the density perturbation $\delta n_{\rm max}/n_0={\rm max}~\left[\delta n(\vec r)\right]/n_0$, where $\delta n(\vec r)=n(\vec r)-n_0$ is the density deviation from the UEG with density $n_0$.

When $\delta n_{\rm max}/n_0\gtrsim 1$ and sufficiently small perturbation wavelength, we access the regime where standard XC functionals fail due to the strong degree of density inhomogeneity and self-interaction \cite{Moldabekov_JCP_2021}.
In the limit $\delta n_{\rm max}/n_0 \ll 1$, the static density response function is defined only by the properties of the UEG and is independent of $A$. In this case, for the static density response function, $\chi(\vec q)$, we have:
\begin{equation}
    \delta n(\vec r)=2A \cos(\vec q \cdot \vec r) \chi(\vec q).
\end{equation}

Within linear response theory, the density response function $\chi(\vec q, \omega)$ can be represented in terms of the ideal response function $\chi_0(\vec q, \omega)$ and the XC kernel,
\begin{eqnarray}\label{eq:kernel}\label{eq:chi}
 \chi(\mathbf{q},\omega) = \frac{\chi_0(\mathbf{q},\omega)}{1 - \left[v(q)
 +K_\textnormal{xc}(\mathbf{q},\omega)\right]\chi_0(\mathbf{q},\omega)}\ ,
\end{eqnarray}
where, for the UEG, $\chi_0(q, \omega)$ is the temperature-dependent Lindhard function.

Following Ref. \cite{dft_kernel}, one can invert Eq. (\ref{eq:chi}) to compute the static XC kernel using   $\chi(\vec q)=\chi(\vec q, \omega=0)$ and  $\chi_0(\vec q)=\chi_0(\vec q, \omega=0)$,
\begin{eqnarray}\label{eq:invert}
 K_\textnormal{xc}(\mathbf{q}) &=& -\left\{
 v(q) + \left( \frac{1}{\chi(\mathbf{q})} - \frac{1}{\chi_0(\mathbf{q})} \right)
 \right\}\ , \\\nonumber
 &=& \frac{1}{\chi_\textnormal{RPA}(\mathbf{q})} - \frac{1}{\chi(\mathbf{q})}\ ,
\end{eqnarray}
where $\chi_\textnormal{RPA}(\mathbf{q})$ is the density response function in the random phase approximation (RPA) defined by setting $K_\textnormal{xc}(\mathbf{q})\equiv0$. In the static case, $\omega=0$, from Eq. (\ref{eq:chi}) we get:
\begin{eqnarray}\label{eq:RPA}
 \chi_\textnormal{RPA}(\mathbf{q}) = \frac{\chi_0(\mathbf{q})}{1 - v(q)\chi_0(\mathbf{q})}\ .
\end{eqnarray}

The XC kernel constitutes the key input to linear-response time-dependent density functional theory.  Clearly, the XC kernels from the hybrid XC functionals are also interesting in their own right since they provide insight into the XC features at different wavenumbers.
We stress that the static XC kernels obtained from hybrid XC functionals for the UEG had not been reported prior to our work; neither for the ground state nor for WDM.

For the UEG, the XC kernel is equivalent to the local field correction (LFC) commonly used in the quantum theory of the electron liquid \cite{giuliani2005quantum, kugler1}.
The relation between the static XC kernel and the static LFC reads
\begin{eqnarray}\label{eq:LFC}
 G(\mathbf{q},\omega) = - \frac{1}{v(q)} K_\textnormal{xc}(\mathbf{q},\omega)\ .
\end{eqnarray}

For a correct description of the UEG and related systems like metals at ambient conditions and at WDM parameters, the LFC must obey the long wave-length expansion based on the compressibility sum-rule ~\cite{IIT_1987},
\begin{eqnarray}\label{eq:CSR}
\lim_{q\to0}G(q) = - \frac{q^2}{4\pi} \frac{\partial^2}{\partial n_0^2} \left( n_0 F_\textnormal{xc} \right)\ ,
\end{eqnarray}
with $n_0=N/V$ denoting the average number density of the electrons in the UEG or the mean value of the conduction electron density  in metals.

We use available exact quantum Monte-Carlo data for $G(q)$ of the UEG to analyze various XC kernels \cite{ PhysRevLett.75.689,  dornheim_ML, Chen2019, PhysRevB.106.L081126, Dornheim_PRL_2020_ESA, Dornheim_PRB_2021}.

Beyond linear response theory, accurate data for the XC kernel (LFC) in combination with ideal density response functions allows one to compute higher order quadratic and cubic density response functions \cite{Dornheim_PRR_2021,Dornheim_JPSJ_2021,Moldabekov_JCTC_2022}.

\section{Simulation Setup}\label{s:setup}

Commonly used parameters for the description of the electrons in the WDM regime are the coupling parameter $r_s$ and the reduced temperature $\theta=T/T_F$, where the former is the mean inter-particle distance in Hartree atomic units and the latter is the ratio of the temperature to the Fermi temperature~\cite{Ott2018}.
The WDM state is defined by the regime where electrons are non-ideal $r_s\gtrsim 1$, and thermal excitations and quantum degeneracy are simultaneously important, i.e., $\theta\sim 1$. Often in experiments, the WDM state is generated by shock-compression and (isochoric) heating of solid-state samples \cite{Denoeud,  Booth2015, PhysRevResearch.2.033366}.
In this regard, understanding the WDM state around metallic densities is particularly important. Thus, we consider the characteristic metallic density of $r_s=2$ and the partially degenerate case with $\theta=1$. This corresponds to an electron temperature $T\simeq 12.5~{\rm eV}$. In the case of a weak perturbation $\delta n_{\rm max}/n_0\ll1$, we additionally consider the ground state by setting $\theta=0.01$. For this case, we analyze the quality of the considered hybrid XC functionals in the UEG limit in terms of quantum Monte-Carlo data for the density response function by Moroni \textit{et al.} \cite{PhysRevLett.75.689}. Furthermore, the analysis of the ground state allows us to gauge the impact of thermal excitation effects on the performance of the hybrid XC functionals at $T\simeq T_F$.

For the perturbation in Eq. (\ref{eq:H}), we consider $A=0.01$, $A=0.1$, $A=0.5$, and $A=1.0$ (in Hartree). These parameters correspond to various degrees of the density inhomogeneity from  $\delta n_{\rm max}/n_0 \lesssim 10^{-2}$ (at $A=0.01$) to $\delta n_{\rm max}/n_0\gtrsim 1$ (at $A=0.5$ and $A=1.0$).
Thus, $A=0.01$ allows one to probe the properties of the system in the UEG limit. We particularly extract the static density response function and the static XC kernel following the approach introduced in Ref. \cite{dft_kernel} and compare them with the PIMC data.

The wavenumber of the perturbation $\vec q$ is directed along the $z$-axis. Note that we drop the vector notation in the following.
The values of the perturbation wavenumber are defined by the length of the simulation cell as $q=2\pi j/L$, where $j\geq 1$ is a positive integer number.
The smallest value of $q$ corresponding to $j=1$ is denoted as $q_{\rm min}$. We consider $q$ in the range from $q_{\rm min}$ to  $4q_{\rm min}$, and up to $5q_{\rm min}$ at $A=0.01$ for the static density response and XC kernel calculations.

The KS-DFT calculations were performed using the ABINIT package \cite{Gonze2020, Romero2020, Gonze2016, Gonze2009, Gonze2005, Gonze2002}.
The calculations at $\theta=1$ were performed with the cubic main cell length $L=7.7703~{\rm Bohr}$, $N_b=200$ orbitals, and the maximal kinetic energy cut-off $13~{\rm Ha}$.
The used cell length corresponds to $N=L^3\times n_0=14$ electrons, where at $r_s=2$ we have $n_0\simeq 0.02984 ~{\rm Bohr^{-3}}$.
The smallest perturbation wavenumber for $14$ particles is $q_{\rm min}[N=14]\simeq 0.84268~q_F$, where $q_F$ is the Fermi wavenumber.
Periodic boundary conditions and a {\textit k}-point grid of size $8\times 8\times 8$ were used.
Self-consistent-field cycles for the solution of the Kohn-Sham equations were converged with absolute differences of the total energy $\delta E<10^{-7}~{\rm Ha}$.
A large number of bands $N_b\gg N$ is needed due to significant thermal smearing of the occupation numbers. At the considered parameters, $N_b=200$ corresponds to the smallest occupation number $f_{\rm min}<10^{-4}$, which is sufficiently small to acquire convergence \cite{Moldabekov_JCP_2021}.
In prior works it was shown that, in the case of the harmonically perturbed electron gas, the calculation results with $N=14$ particles are not affected by finite size effects \cite{Moldabekov_JCP_2021, Moldabekov2022, dft_kernel, Dornheim_2020, doi:10.1063/1.5123013, DVB2020:nonlinear}. In this work, keeping other parameters equivalent, the convergence is cross-checked by comparing the data for $N=14$ to the results computed with $N=20$ and $N_b=300$ at $q=2q_{\rm min}$ (with $A=0.01$).
Additionally, we used three different particle numbers in the simulation cell for the ground state calculations:  $N=38$, $54$, and $66$; with {\textit k}-point grid of $8\times 8\times 8$. The agreement and consistency of the results for all considered hybrid XC functionals computed with different particle numbers reassure the convergent character of the calculations.
Finally, for both $\theta=1$ and $\theta=0.01$, we performed KS-DFT calculations with the XC functional set to zero yielding the static density response function of the UEG in the random phase approximation (RPA) (see Sec. \ref{s:t0_ueg}) and compared it to the analytical Lindhard function (in $N\to \infty$ limit).
The perfect agreement of our KS-DFT data with the analytical Lindhard function (shown in Figs. \ref{fig:chi_t0}a and \ref{fig:chi_t1}a) substantiates our confidence in the convergence of our calculations.

We report results for a total of $155$ KS-DFT calculations for the HSE06, HSE03, PBE0, PBE0-1/3, and B3LYP hybrid XC functionals with different choices of parameters (amplitude $A$ and wavenumber $q$) in the density perturbation.

A fraction of the PIMC data (with $q=q_{\rm min},~2q_{\rm min} $ and $3q_{\rm min}$) used in this work has been reported in the analysis of the ground state LDA, PBE, PBEsol, AM05, and SCAN functionals in Ref. \cite{Moldabekov_JCP_2021}. It was shown there that the quality of the KS-DFT calculations based on these XC functionals has the tendency to deteriorate with an increasing wavenumber of the perturbation.
Here, we test the hybrid XC functionals in the regime where the functionals in the previous analysis have failed.  We have therefore generated new PIMC data for $q=4q_{\rm min}$ at $A=0.1, ~0.5$, and $1.0$.

Since a detailed introduction to the direct PIMC method~\cite{cep} and the associated fermion sign problem~\cite{dornheim_sign_problem} have been presented elsewhere, we here restrict ourselves to a brief summary of the main simulation details. We have used the extended ensemble approach introduced in Ref.~\cite{Dornheim_PRB_nk_2021}, which is a canonical adaption of the worm algorithm by Boninsegni \emph{et al.}~\cite{boninsegni1,boninsegni2}. Moreover, we use $P=200$ imaginary-time propagators from a primitive factorization scheme, which is sufficient to ensure convergence; see the Supplemental Material of Ref.~\cite{PhysRevLett.125.085001} for a corresponding analysis.

\section{Results}\label{s:results}

We start with the weakly perturbed electron gas where $|\delta n(\vec r)|/n_0\ll 1$ and consider both the ground state and the parameters relevant for WDM.
We analyze the results both in terms of the static density response function $\chi(q)$ and the static XC kernel $K_{\rm xc}(q)$.
In the next step, we consider strongly perturbed systems with $|\delta n(\vec r)|/n_0\gtrsim 1$ and provide the analysis of the accuracy of the electron density calculations based on the considered hybrid XC functionals.

\subsection{Weakly perturbed electron gas}\label{s:weak}

\subsubsection{Ground state results}

Let us start from the ground state with $\theta=0.01$.
First of all, the analysis of the $T\to 0$ limit helps to establish a baseline for further analyzing the performance of the hybrid XC kernels at high temperatures.
Second, for the construction of the LDA and GGA (e.g. PBE) XC functionals, the reproduction of the LFC of the UEG at $q<2q_F$  was one of the main criteria for an adequate description of metals \cite{PBE}. Thus, it is interesting to see to what degree the considered hybrid XC functionals can reproduce the LFC of the UEG.

\begin{figure}
\center
\includegraphics[width=0.45\textwidth]{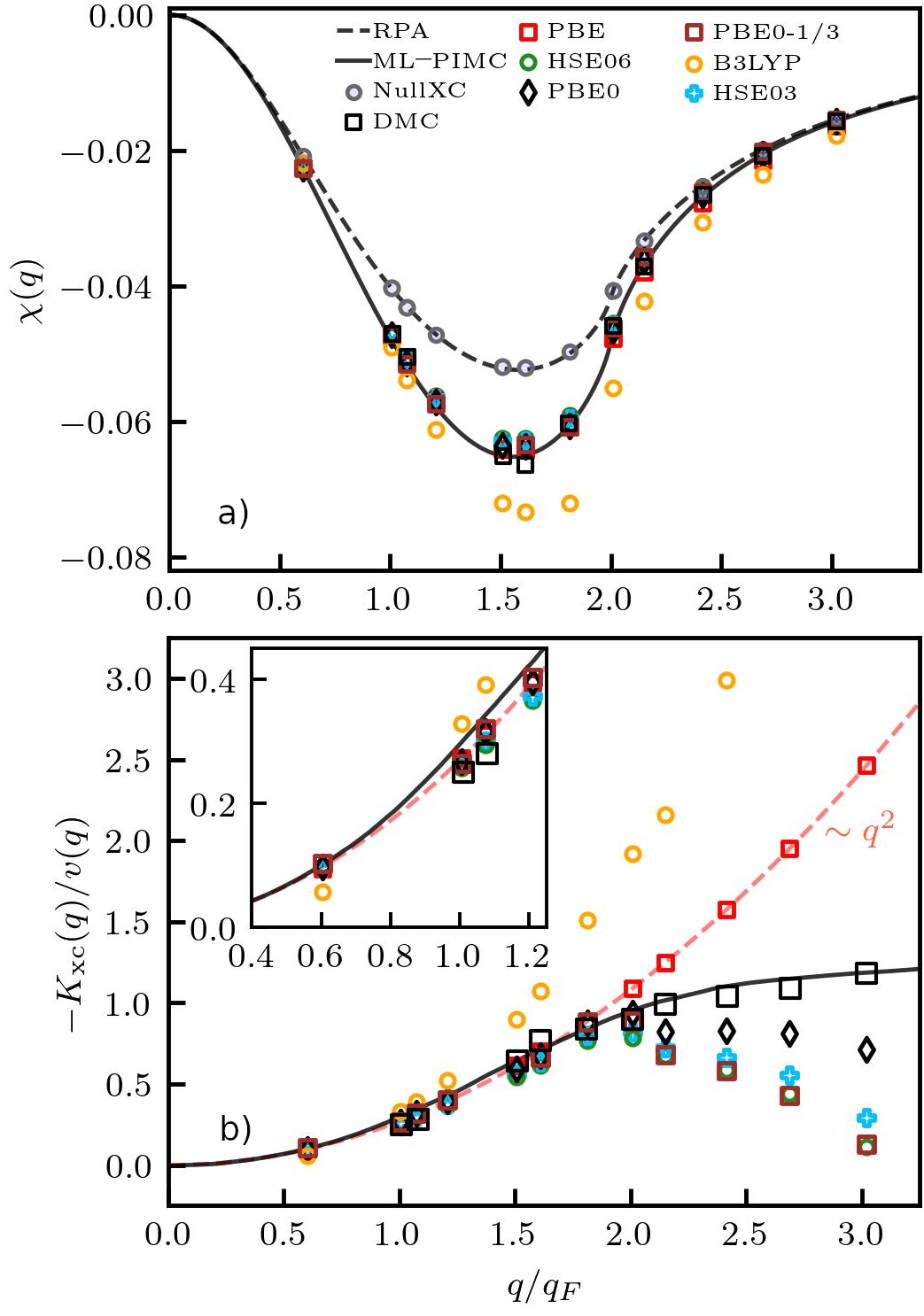}
\caption{ \label{fig:chi_t0}
The linear static electron density response function (top) and the static XC kernel (bottom) at $\theta=0.01$ and $r_s=2$. Solid black line: exact UEG results based on the neural-net representation of Ref.~\cite{dornheim_ML}. Black squares: exact QMC results by Moroni \emph{et al.}~\cite{PhysRevLett.75.689}. The RPA result for the density response function in the top panel is given by the dashed line. The grey circles in the top panel are the KS-DFT results for the density response function with the XC functional set to zero (denoted as NullXC). The other symbols distinguish KS-DFT results computed using different XC-functionals.
}
\end{figure}

The results for the static density response function and the static XC kernel are shown in Fig. \ref{fig:chi_t0}a) and Fig. \ref{fig:chi_t0}b), respectively.
In Fig. \ref{fig:chi_t0}, we compare the KS-DFT results  with the diffusion quantum Monte Carlo (DMC) data by Moroni \textit{et al.} (black squares) \cite{PhysRevLett.75.689} computed for $T\to 0$.  In addition, we show the results from the machine-learning representation (ML-PIMC) by Dornheim {\textit {et al.}} of extensive PIMC data at finite temperatures \cite{dornheim_ML} and the DMC data  \cite{PhysRevLett.75.689} for $T=0$. Besides the results based on the hybrid XC functionals, in Fig. \ref{fig:chi_t0} we also show results computed using PBE. The latter comparison is meaningful because PBE0, PBE0-1/3, HSE06, and HSE03 are based on PBE.

Before discussing our results on the hybrid XC functionals, we validate the absence of any significant finite size effects in Fig. \ref{fig:chi_t0}a by comparing the analytical Lindhard function with the static density response from the KS-DFT calculations where the XC functional is set to zero (labeled NullXC \cite{dft_kernel}). The latter is equivalent to the RPA and does indeed agree with the analytical Lindhard function as expected.

In Fig. \ref{fig:chi_t0}a),  we observe that the results for $\chi(q)$ based on PBE0, PBE0-1/3, HSE06, HSE03, and PBE are closely grouped and are in rather good agreement with the reference DMC data. In contrast, the B3LYP results for $\chi(q)$ significantly deviate from the reference ML-PIMC data at $q>q_F$. This is expected since the B3LYP does not attain the UEG limit by design. This was elucidated by Paier \textit{et al.} for the unperturbed UEG \cite{B812838C}.
Here we further substantiate it for finite perturbation wavenumbers in the range $0.6\leq q \leq 3q_F$ as it can be seen in Fig. \ref{fig:chi_t0}b) for the static XC kernel, where a strong deviation of the B3LYP results from the DMC and ML-PIMC data is clearly visible. Furthermore, from Fig. \ref{fig:chi_t0}b, we see that the B3LYP data does not obey the compressibility sum-rule in Eq. (\ref{eq:CSR}) with quadratic $q$ dependence of the LFC at $q\lesssim q_F$, which is a crucial criterion for the description of metals \cite{PBE}; this can be seen particularly well in the inset that shows a magnified segment.

In contrast to B3LYP, the LFCs of PBE0, PBE0-1/3, HSE06, and HSE03 reproduce the quadratic dependence on the wavenumber at $q\lesssim 1.5 q_F$ in accordance with Eq. (\ref{eq:CSR}), as shown in Fig. \ref{fig:chi_t0}b).
We also observe in Fig. \ref{fig:chi_t0}b) that the PBE results have a quadratic dependence on $q$ at all considered wave numbers since it was designed in a way that the gradient corrections to the exchange and to the correlation parts exactly cancel each other in the limit of the UEG \cite{PBE}.

At large wavenumbers, we observe from Fig. \ref{fig:chi_t0}b) that the PBE0, PBE0-1/3, HSE06, and HSE03 results deviate from the DMC data at $q\gtrsim 2 q_F$. Nevertheless, at $2 q_F\leq q\leq 3q_F$, the results from the considered hybrid XC functionals exhibit smaller absolute deviations from the DMC data compared to the PBE results.
An interesting feature of the LFC based on the PBE0, PBE0-1/3, HSE06, and HSE03 is that it has a maximum somewhere around $2q_F$ and a monotonically decreasing trend with increasing $q$ at $q>2 q_F$. A similar behavior---with a somewhat worse agreement with the DMC data---was reported by Moldabekov {\textit et al.} for the AM05 \cite{PhysRevB.72.085108} functional, which is a semi-local GGA. Another related observation is that AM05 provides as accurate a description as the PBE0 and HSE06 hybrid XC functionals for lattice constants, bulk moduli, and surfaces of solids \cite{Mattsson}.
The latter property, i.e., the electronic behavior at surfaces, corresponds to a domain of strongly inhomogeneous electron
densities characterized by large wave numbers $q\gg q_F$ \cite{PhysRevB.104.045118}. Thus, it is an interesting observation that the AM05 XC kernel has a similar behavior as the XC kernel computed using the HSE06, HSE03, PBE0, and PBE0-1/3 functionals at $q>2q_F$.
We do not further extend the analysis of the observed similarities between the AM05 XC kernel and the hybrid XC kernels since it would deviate from the main focus of this work, but we leave such investigation as a possible direction for future research.

\begin{figure}
\center
\includegraphics[width=0.45\textwidth]{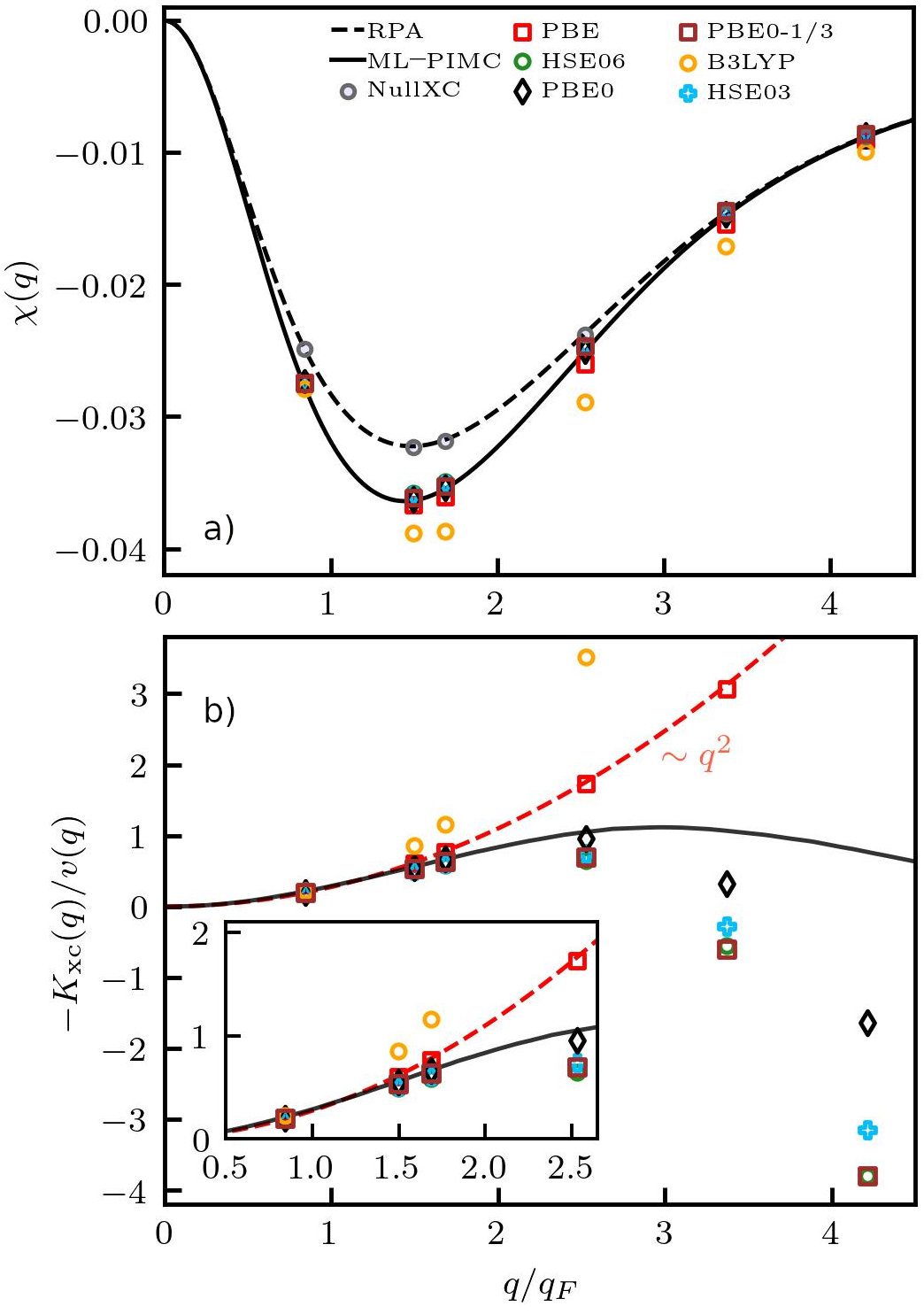}
\caption{  \label{fig:chi_t1}
The linear static electron density response function (top) and the static XC kernel (bottom) at $\theta=1$  and $r_s=2$.
Solid black line: exact UEG results based on the neural-net representation of Ref.~\cite{dornheim_ML}. The RPA result for the density response function in the top panel is given by the dashed line. The grey circles in the top panel are the KS-DFT results for the density response function with the XC functional set to zero (denoted as NullXC). The other symbols distinguish KS-DFT results computed using different XC-functionals.
}
\end{figure}

In Fig. \ref{fig:chi_t0}b), one can see that at $q\lesssim 1.5 q_F$, the PBE0, PBE0-1/3, HSE06, HSE03, and PBE functionals provide static XC kernels in excellent agreement with the DMC data.  Additionally, the PBE0 and PBE0-1/3 results are much closer to the DMC data at $q\simeq 2q_F$ compared to HSE06, HSE06, and PBE. At $q\gtrsim 2q_F$, PBE0 provides a much better description of the static XC kernel of the UEG compared to the other hybrid XC functionals.

In addition, we find that increasing the contribution of HF exchange from $1/4$ (as in PBE0) to $1/3$ (as in PBE0-1/3) lowers the resulting values of $K_{\rm xc}(q)$ at $q>2q_F$.
Keeping the coefficient $1/4$ in front of the HF exchange fixed in Eq. (\ref{eq:pbe0}), the same effect is achieved by using a screened potential instead of the Coulomb potential as illustrated by the static XC kernel based on the HSE06 and the HSE03. On the other hand, the behavior of the $K_{\rm xc}(q)$ at $q<2q_F$ is not sensitive to the screening of the Coulomb potential or the variation of the HF exchange contribution. One could adjust the screening parameter or the weight of the HF exchange contribution in order to accurately reproduce the exact data for the $K_{\rm xc}(q)$ in a wider range of wavenumbers.

The presented results clearly illustrate the utility of this assessment in terms of a harmonic perturbation \cite{dft_kernel} as a device for understanding and developing improved XC functionals.

\subsubsection{Results at a high temperature $T\simeq T_F$}
Next, we analyze the partially degenerate case with $\theta=1$.
Our new results for the static density response function and the static XC kernel of the UEG at $\theta=1$ are presented in Fig. \ref{fig:chi_t1}a) and Fig. \ref{fig:chi_t1}b), respectively.
Similar to the ground state, we observe that the B3LYP results for the static density response function (Fig. \ref{fig:chi_t1}a) and the static XC kernel (Fig. \ref{fig:chi_t1}b) disagree significantly with the reference ML-PIMC data at $q\gtrsim q_F$. Also, the B3LYP data for the static XC kernel does not obey a quadratic dependence on the wavenumber at $q\lesssim q_F$ and approaches the PBE result from above with a decrease in $q$ and crosses the PBE data at $q\simeq 0.84 q_F$. The PBE data show a quadratic dependence on $q$ in the entire considered range of wavenumbers and agree well with the reference data (ML-PIMC) at $q\lesssim 1.5 q_F$. The PBE0, PBE0-1/3, HSE06, and HSE03 results are in close agreement with the reference data (ML-PIMC) at $q\lesssim 2.5 q_F$ (with the PBE0 being the most accurate).
Therefore, these functionals lead to a better description of the density response compared to PBE.
At $q\gtrsim 2.5 q_F$, the PBE0, PBE0-1/3, HSE06, and HSE03 functionals produce static XC kernels that decrease with increasing $q$ and eventually become negative at $q\gtrsim 3.5 q_F$. A similar behavior has been reported for the AM05 functional at $T\simeq T_F$ \cite{dft_kernel}.
In general, the trends in the $q$-dependence and main features of $\chi(q)$ and $K_{\rm xc}(q)$ of the considered hybrid XC functionals are similar to those in the ground state discussed above. This indicates that the partial occupation of the orbitals at high temperatures does not lead to fundamentally new features in the hybrid XC functional results.

With that, we conclude our consideration of the weak perturbation limit. We continue further with the case of the strong density inhomogeneity.

\begin{figure}
\center
\includegraphics[width=0.45\textwidth]{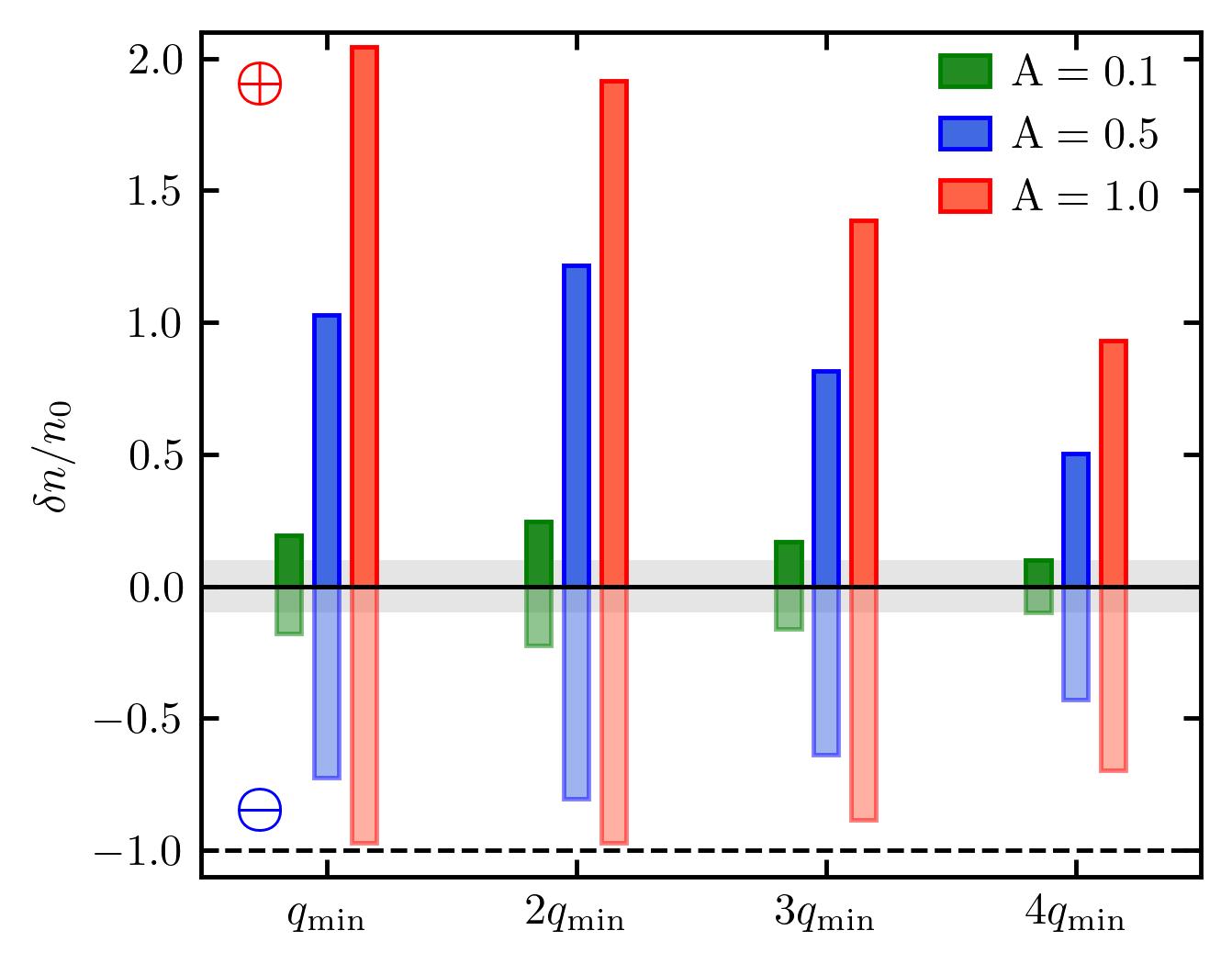}
\caption{ \label{fig:max_den}
The largest value of the density perturbation at different external field amplitudes and wavenumbers in the density accumulation region denoted as $\oplus$ (positive values) and the density depletion region denoted as $\ominus$ (negative values). The smallest considered value of the perturbation wavenumber is $q\simeq 0.84 q_F$. The shaded area indicates the density perturbations  $\delta n\leq 0.1 n_0$. The dashed horizontal line represents lower bound for the density perturbation in the depletion region since $n=n_0+\delta n$ cannot be negative. The data is from the PIMC calculations at $r_s=2$ and $\theta=1$.
}
\end{figure}
\begin{figure}[ht!]
\center
\includegraphics[width=0.45\textwidth]{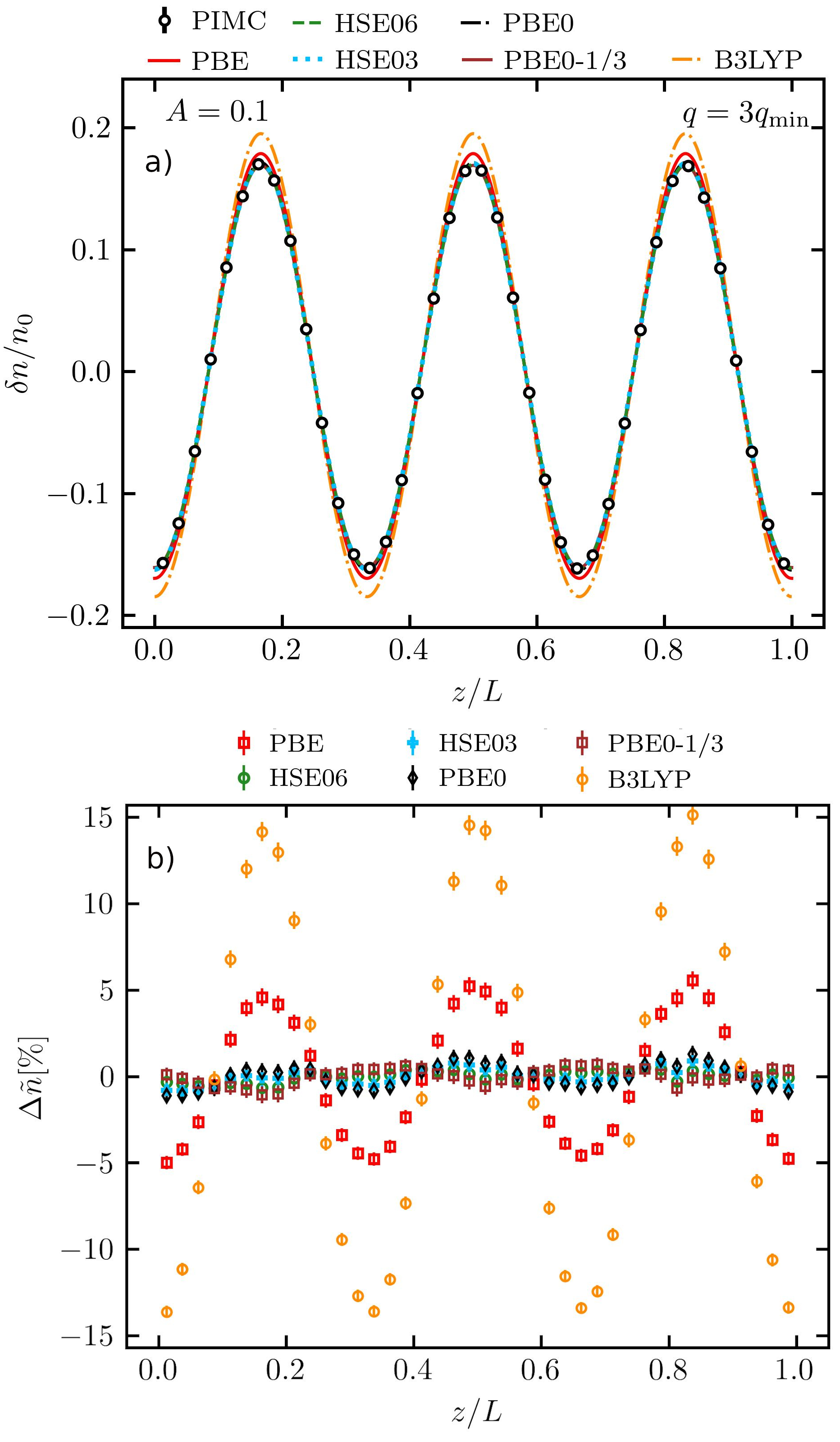}
\caption{ \label{fig:3q_A01}
a) The density perturbation $\delta n(\vec r)$  and b) the density error $\Delta \widetilde n ~[\%]$ profiles along the perturbation direction at $r_s=2$ and $\theta=1$.
}
\end{figure}

\subsection{Strongly inhomogeneous electron gas}\label{s:inhom}

To generate a strong density perturbation, we apply the external harmonic perturbation as defined in Eq.~(\ref{eq:H}) with amplitudes $A=0.1,~0.5$ and $1.0$ and wavenumbers in the range from $q=q_{\rm min}\simeq 0.84~q_F$ up to $q=4q_{\rm min}$. The amplitude of the density inhomogeneity is qualitatively illustrated in Fig. \ref{fig:max_den}, where we plot the largest values of the density accumulation ($\delta n_{\oplus}$) and of the density depletion ($\delta n_{\ominus}$) in the simulation cell due to the external harmonic field with different wavenumbers.
The gray area indicates density perturbations less than $0.1n_0$.
We see that at the considered amplitudes of the perturbation we always have $|\delta n_{\oplus}|\geq 0.1 n_0$ and $|\delta n_{\ominus}|\geq 0.1 n_0$. These density perturbation magnitudes are well beyond the linear response regime considered in Sec. \ref{s:weak}.
At $A=0.5$, we have $|\delta n_{\oplus}|\sim n_0$ and $|\delta n_{\ominus}|\sim 0.5 n_0$.
Furthermore, at $A=1.0$, the electron density values almost vanish in the density depletion regions with $\delta n_{\ominus}\to -n_0$ at $q=q_{\rm min}$ and $q=2q_{\rm min}$, where a physical lower bound $\delta n_{\ominus}>-n_0$ is in place (indicated by the dashed horizontal line in Fig. \ref{fig:max_den}). Correspondingly,  most of the electrons are localized in the density accumulation regions with  $\delta n_{\oplus}\approx 2n_0$.

An important point for our analysis is that at fixed values of $A$, the density perturbation magnitude can have a non-monotonic dependence on $q$ due to a stronger response of electrons around $q=2q_{\rm min}\approx 1.5q_F$ as it is clearly illustrated in Fig. \ref{fig:max_den} for $A=0.1$ and $A=0.5$. A stronger density response at the wavelength corresponding to the mean inter-particle distance $d\simeq 2r_s$, i.e at $q\simeq 2\pi/d\simeq 2q_F$, is a manifestation of the fact that it is energetically more favorable for particles to be localized at the distance $r \simeq d$ from each other. This was demonstrated and used for the explanation of the roton feature in the excitation spectrum of the UEG by Dornheim \textit{et al.,} \cite{https://doi.org/10.48550/arxiv.2203.12288}. Contrarily, at fixed values of $q$, the increase in $A$ always leads to a monotonic increase in the density perturbation amplitude.
Therefore, to analyze the error of the hybrid XC functionals in the KS-DFT calculations, it is more transparent to compare the deviations from the exact PIMC data for various $A$ values at a fixed $q$, rather than vice versa.

We perform an error analysis following Refs. \cite{Moldabekov_JCP_2021, Moldabekov_PRB_2022}, where we use the relative density deviation between the KS-DFT data and the reference QMC data defined as
\begin{equation}\label{eq:Dn}
    \Delta \widetilde n ~[\%]=\frac{\delta n_{\rm DFT}-\delta n_{\rm QMC}}{{\rm max}\{\delta n_{\rm QMC}\}} \times 100,
\end{equation}
where ${\rm max}\{\delta n_{\rm QMC}\}$ is the maximum deviation of the QMC data from $n_0$.
From a physical perspective, Eq. (\ref{eq:Dn}) defines an error in the density response computed using KS-DFT in comparison with the exact density response from the PIMC calculations. It becomes obvious by considering a weak perturbation where $\delta n(r)=2A\cos(qr)\chi(q)$, and Eq. (\ref{eq:Dn}) reduces to the error evaluation of the linear density response function \cite{Dornheim_2020_ppcf}.

Before we dive into the details of the error analysis of the hybrid XC functionals, in Fig. \ref{fig:3q_A01} we illustrate the density perturbation $\delta n(\vec r)=n(\vec r)-n_0$ and the density error $\Delta \widetilde n$ profiles when the harmonic perturbation with $A=0.1$ and $q=3q_{\rm min}$ is applied.
From Fig. \ref{fig:3q_A01}a), we observe that the PBE data exhibits a certain deviation from the PIMC data at the maximum and minimum values of $\delta n(\vec r)$. The B3LYP results show a significant disagreement with the PIMC data in both the density accumulation region (with $\delta n(\vec r)>1$) and the density depletion region (with $\delta n(\vec r)<1$).
The results for the other considered hybrid XC functionals are in a close agreement with the exact PIMC data.
This is further reaffirmed by $\Delta \widetilde n$ in Fig. \ref{fig:3q_A01}b, where $|\Delta \widetilde n|\lesssim 1\%$ for the PBE0, PBE0-1/3, HSE06, and HSE03 results.
Additionally, from Fig. \ref{fig:3q_A01}b) we clearly see that the PBE (the B3LYP) data has an error of up to $\pm 5\%$ ($\pm 15\%$) in the relative density deviation $\Delta \widetilde n$. Obviously, in this case, the PBE0, PBE0-1/3, HSE06, and HSE03 hybrid XC functionals provide a significant improvement due to mixing in a fraction of HF exchange.

To perform a general analysis of the errors in the density response, we consider the largest value of the $\Delta \widetilde n$ in the density accumulation region, $\Delta \widetilde n_{\oplus}$, and the largest magnitude of  the $\Delta \widetilde n$ in the density depletion region, $\Delta \widetilde n_{\ominus}$, for different $A$ and $q$.

\begin{figure}
\center
\includegraphics[width=0.45\textwidth]{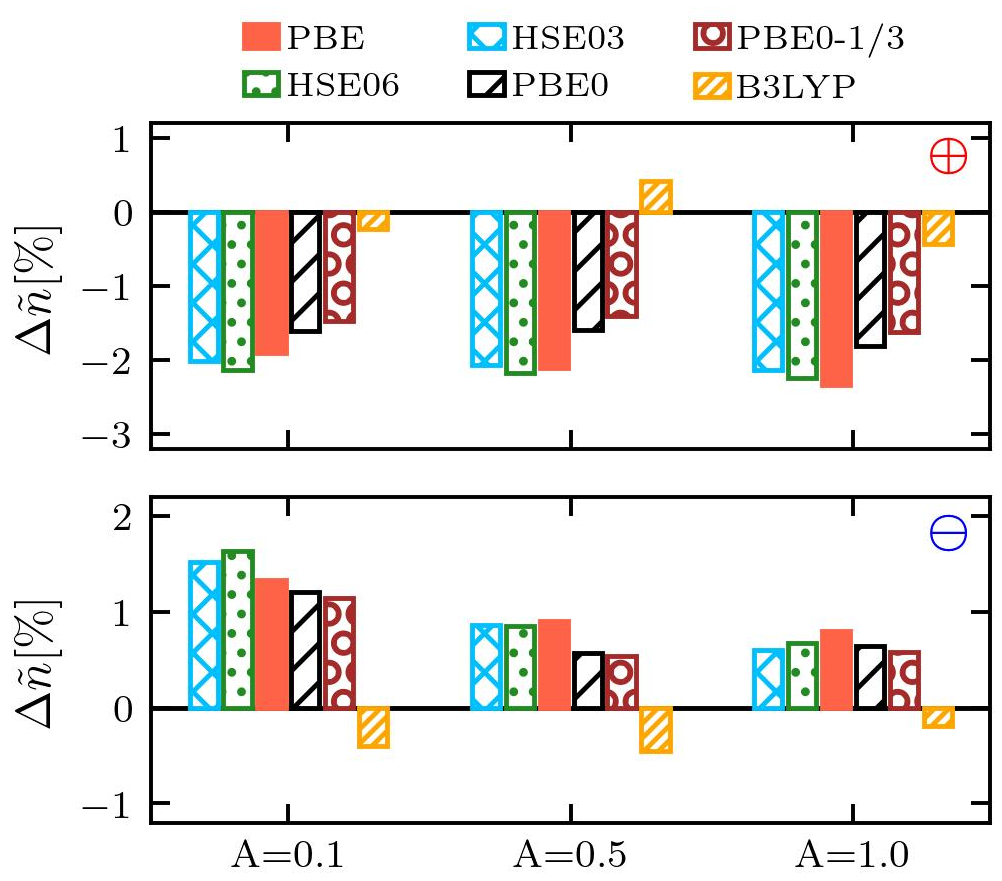}
\caption{ \label{fig:bar1}
Histogram representation of the largest value of the error in the relative density deviation of the KS-DFT data from the exact PIMC data, Eq. (\ref{eq:Dn}), in the density accumulation region (the top panel denoted as $\oplus$) and in the density depletion region (the bottom panel denoted as $\ominus$) for different $A$ at $q=q_{\rm min}\simeq 0.84~ q_F$.
The results are computed for $r_s=2$ and $\theta=1$.
}
\end{figure}

\begin{figure}
\center
\includegraphics[width=0.45\textwidth]{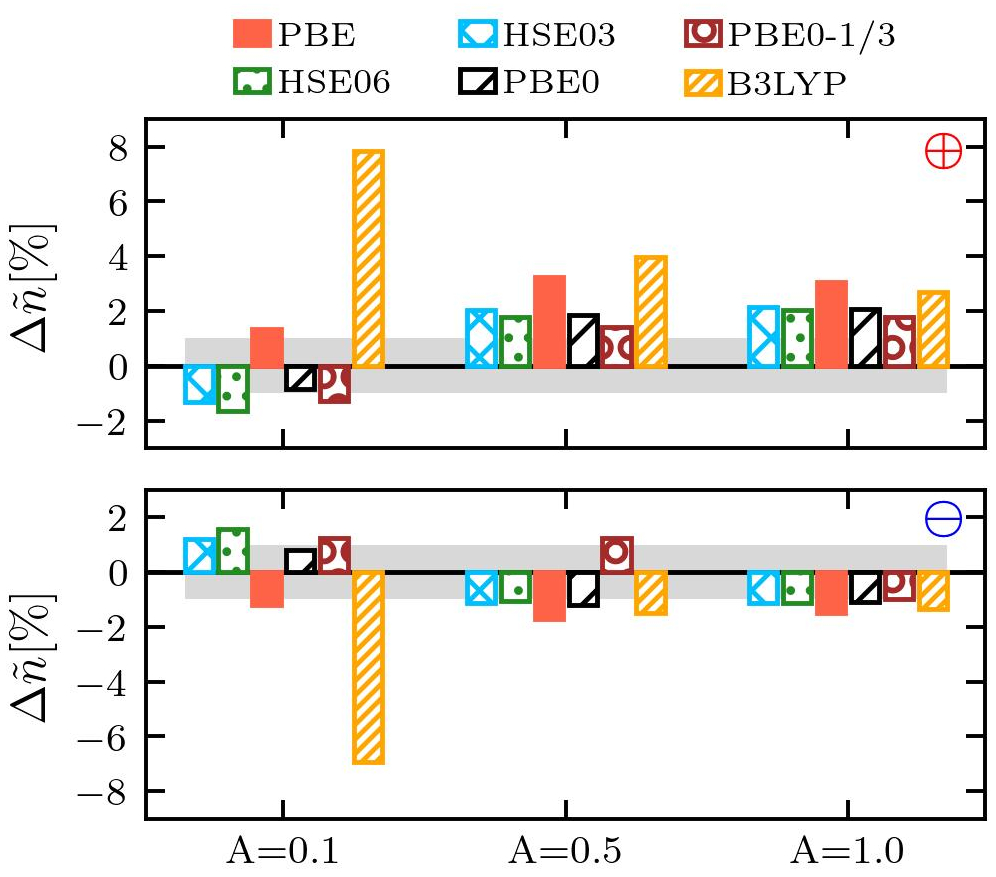}
\caption{ \label{fig:bar2}
Histogram representation of the largest value of the error in the relative density deviation of the KS-DFT data from the exact PIMC data, Eq. (\ref{eq:Dn}), in the density accumulation region (the top panel denoted as $\oplus$) and in the density depletion region (the bottom panel denoted as $\ominus$) for different $A$ at $q=2q_{\rm min}\simeq 1.685~ q_F$.
The grey area corresponds $\Delta \widetilde n_{\oplus}\leq \pm 1~\%$ and $\Delta \widetilde n_{\ominus}\leq \pm 1~\%$.
The results are computed for $r_s=2$ and $\theta=1$.
}
\end{figure}

\begin{figure}
\center
\includegraphics[width=0.45\textwidth]{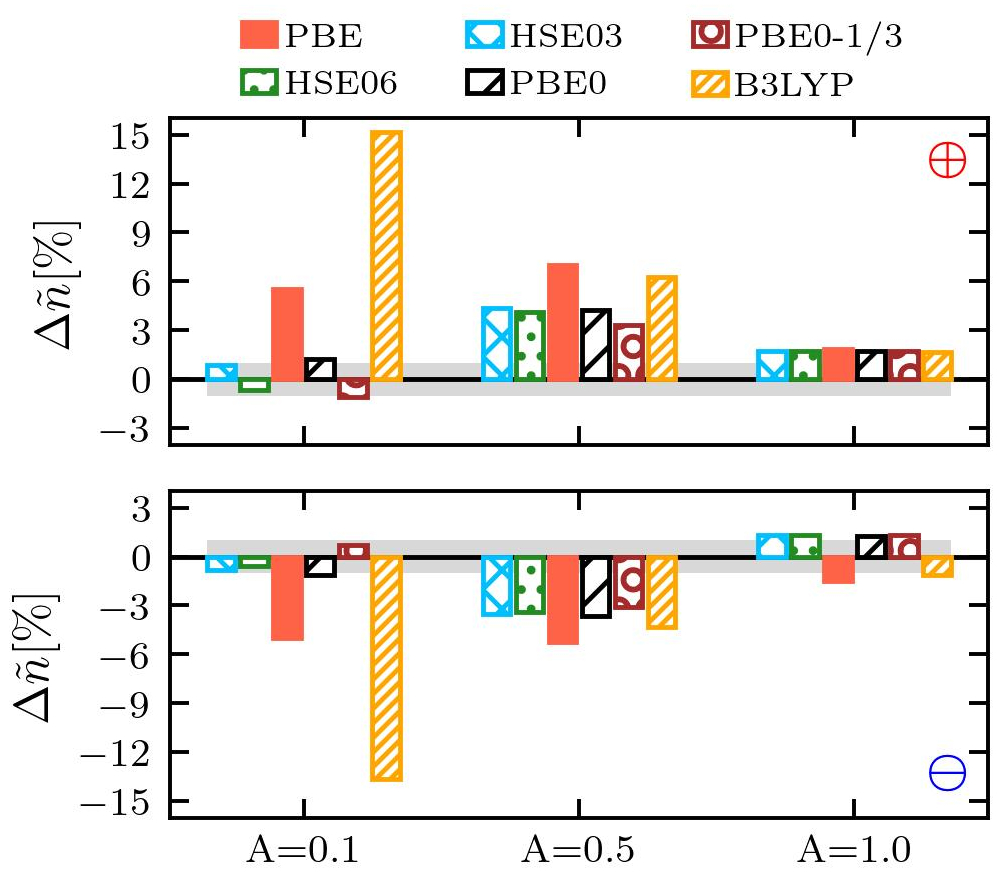}
\caption{ \label{fig:bar3}
Histogram representation of the largest value of the error in the relative density deviation of the KS-DFT data from the exact PIMC data, Eq. (\ref{eq:Dn}), in the density accumulation region (the top panel denoted as $\oplus$) and in the density depletion region (the bottom panel denoted as $\ominus$) for different $A$ at $q=3q_{\rm min}\simeq 2.528~ q_F$.
The grey area corresponds $\Delta \widetilde n_{\oplus}\leq \pm 1~\%$ and $\Delta \widetilde n_{\ominus}\leq \pm 1~\%$.
The results are computed for $r_s=2$ and $\theta=1$.
}
\end{figure}

\begin{figure}
\center
\includegraphics[width=0.45\textwidth]{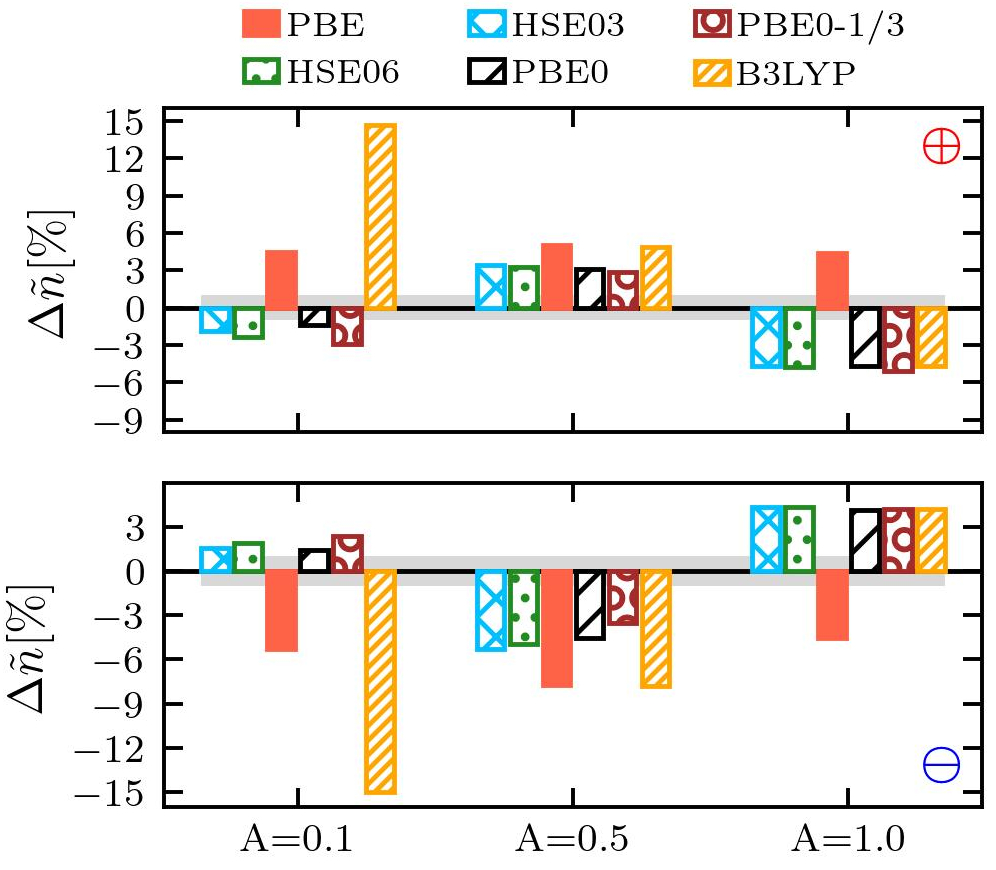}
\caption{ \label{fig:bar4}
Histogram representation of the largest value of the error in the relative density deviation of the KS-DFT data from the exact PIMC data, Eq. (\ref{eq:Dn}), in the density accumulation region (the top panel denoted as $\oplus$) and in the density depletion region (the bottom panel denoted as $\ominus$) for different $A$ at $q=4q_{\rm min}\simeq 3.371~ q_F$.
The grey area corresponds $\Delta \widetilde n_{\oplus}\leq \pm 1~\%$ and $\Delta \widetilde n_{\ominus}\leq \pm 1~\%$.
The results are computed for $r_s=2$ and $\theta=1$.
}
\end{figure}

In Fig. \ref{fig:bar1}, we show the histogram representation of the largest value of the error in the relative density deviation as defined by $\Delta \widetilde n_{\oplus}$ and $\Delta \widetilde n_{\ominus}$ at $q=q_{\rm min}\simeq 0.84~ q_F$ and different values of $A$. We observe that the B3LYP data has the smallest error with $\Delta \widetilde n_{\oplus}<1~\%$ and $\Delta \widetilde n_{\ominus}<1~\%$ compared to other KS-DFT data. However, it appears to be the crossing point between the B3LYP density response and the exact density response as the function of the parameters $A$ and $q$, where one can say that the good performance of B3LYP is due to chance. We have already observed such a crossing point when considering the linear density response function in the case of the weak perturbation in Sec. \ref{s:weak}. Indeed, the B3LYP data significantly deviate from the reference PIMC data at other considered $q>q_{\rm min}$ values at $A=0.1,~0.5,$ and $1.0$ (see Figs. \ref{fig:bar2}-\ref{fig:bar4}). In fact, at $A=0.1$ and $q=2q_{\rm min},~3q_{\rm min},$ and $4q_{\rm min}$, the B3LYP data have the largest errors in both density accumulation and depletion regions among all considered XC functionals.
This is consistent with the conclusions drawn from the linear density response function in Sec. \ref{s:weak}.
The reason is that at $A=0.1$, with $|\delta n_{\oplus}|\sim 0.1 n_0$ and $|\delta n_{\ominus}|\sim 0.1 n_0$ (see Fig. \ref{fig:max_den}), the linear density response function has the largest contribution to the total density response within non-linear response theory, which includes quadratic and cubic density response functions \cite{Dornheim_PRR_2021, Moldabekov_JCTC_2022}. This reasoning is not applicable at $A=0.5$ and $A=1.0$ since the density deviation from the UEG case is well beyond of the regular perturbation expansion, which is valid for  $|\delta n(\vec r)|/n_0\ll 1$.


From Fig. \ref{fig:bar1}, one can see that at $q=q_{\rm min}\simeq 0.84~ q_F$ and $A=0.1,~0.5,$ and $1.0$, PBE0, PBE0-1/3, HSE06, and HSE03 have a similar performance as PBE, with $\Delta \widetilde n_{\oplus}\sim -2~\%$ and $\Delta \widetilde n_{\ominus}\sim ~1\%$. As shown in Fig. \ref{fig:bar2}, at $q=2q_{\rm min}\simeq 1.685~ q_F$, the PBE0, PBE0-1/3,  HSE06, and HSE03 results are a bit more accurate than PBE with the improvement of the accuracy approximately between $1\%$ and  $2\%$ at $A=0.5$ and $A=1.0$ in the density accumulation region

At  $q=3q_{\rm min}\simeq 2.528~ q_F$ and $A=0.1$, we observe the largest improvement in the accuracy of the density response using PBE0, PBE0-1/3, HSE06, and HSE03 compared to PBE (see Fig. \ref{fig:bar3}). At these parameters, the PBE data deviates significantly with about $5\%$ in $\Delta \widetilde n_{\oplus}$ and $\Delta \widetilde n_{\ominus}$. In contrast, the PBE0, PBE0-1/3, HSE06, and HSE03 results are highly accurate with deviations of $|\Delta \widetilde n_{\oplus}|\leq 1\%$ and $|\Delta \widetilde n_{\ominus}|\leq 1\%$.
From Fig. \ref{fig:bar3}, one can see that at $q=3q_{\rm min}$ and $A=0.5$, PBE0, PBE0-1/3, HSE06, and  HSE03 provide an improvement of a few percent in accuracy compared to PBE.
At $A=1.0$ and $q=3q_{\rm min}$, all considered XC functionals show similar accuracy with $|\Delta \widetilde n_{\oplus}|\lesssim 2\%$ and $|\Delta \widetilde n_{\ominus}|\lesssim 2\%$.

Finally, in Fig. \ref{fig:bar4}, we show results for $q=4q_{\rm min}\simeq 3.371~ q_F$.
Similar to the case with  $q=3q_{\rm min}\simeq 2.528~ q_F$,  PBE0, PBE0-1/3, HSE06, and HSE03 yields significantly higher accuracy than PBE for the density response at $A=0.1$ and $A=0.5$, with up to two times smaller errors in $|\Delta \widetilde n_{\oplus}|$ and/or $|\Delta \widetilde n_{\ominus}|$. 
At last, for $A=1.0$ and $q=4q_{\rm min}$, we observe similar accuracy for all considered XC functionals with $|\Delta \widetilde n_{\oplus}|\lesssim 4\%$ and $|\Delta \widetilde n_{\ominus}|\lesssim 4\%$.

\section{Conclusions and Outlook}\label{s:end}

Before drawing general conclusions, we recall that in previous work Moldabekov \textit{et al.} \cite{Moldabekov_JCP_2021} had assessed the accuracy both of the LDA and GGA functionals for the density response in the warm dense electron gas. They had shown that the PBEsol and LDA functionals were approximately as accurate as PBE for the density response in the partially degenerate case with $T\simeq T_F$ for both weak and strong perturbations. The SCAN functional ---being highly accurate in the ground state---had deviated significantly from the reference PIMC data at $q>q_F$ \cite{Moldabekov_JCP_2021, dft_kernel}.

In this work, the benchmark against the exact PIMC data at $q\lesssim 3q_F$ for both weak and strong perturbations demonstrates that the PBE0, PBE0-1/3,  HSE06, and  HSE03 hybrid XC functionals yield a more accurate description of the density response compared to PBE.
In contrast, B3LYP is significantly less accurate for the perturbed electron gas compared to the PBE or the other hybrid functionals.

Furthermore, we presented new results for the static XC kernel computed with the PBE0, PBE0-1/3, HSE06, HSE03, and B3LYP functionals for both the ground state and partially degenerate case with $T\simeq T_F$. Among the considered XC functionals, it has been revealed that PBE0 yields the closest agreement with the exact PIMC and DMC reference data. Particularly, at $q<3q_F$, the PBE0 static XC kernel is practically exact. For the partially degenerate regime, this is a significant improvement over LDA, PBEsol, AM05, and SCAN static XC kernels \cite{dft_kernel}. At $q> 3q_F$ and $T\simeq T_F$, none of the aforementioned LDA, GGA, meta-GGA, and hybrid XC functionals are able to accurately describe the static LFC of the UEG, $G(q)=(q^{2}/4\pi) K_{\rm xc}(q)$. This is likely not a critical point for the density response function since $K_{\rm xc}(q)\ll1$ and $K_{\rm xc}(q)$ do not introduce a significant correction into $\chi(q)$ as evident from Eq. (\ref{eq:chi}).
On the other hand, large wave numbers $q>3q_F$ are important for the accurate description of the electronic static structure factor $S(q)$ and, thus, for the description of the pair electron-electron distribution function $g(r)$ at small distances \cite{Dornheim_PRL_2020_ESA, Dornheim_PRB_2021}. Experimentally, such large wave numbers are probed in WDM and dense plasmas using the XRTS signal collected at large scattering angles \cite{GFGN2016:matter,Preston,Frydrych2020}.

The static XC kernel, being a crucial quantity in linear-response time-dependent density functional theory, enables a unique insight into the performance of the various XC functionals. Thus, in combination with the exact quantum Monte-Carlo data for the UEG \cite{ PhysRevLett.75.689,  dornheim_ML, Chen2019, PhysRevB.106.L081126, Dornheim_PRL_2020_ESA, Dornheim_PRB_2021}, it also can help with the construction of improved XC functionals for WDM applications, where the XRTS signal from large $q$ values is crucial for the diagnostics of system parameters in experiments.

We are convinced that our present findings will be useful for future DFT calculations of WDM, in general, and for the many applications that utilize the XC kernel~\cite{dft_kernel}, in particular.


\section*{Acknowledgments}
This work was funded by the Center for Advanced Systems Understanding (CASUS) which is financed by Germany’s Federal Ministry of Education and Research (BMBF) and by the Saxon state government out of the State budget approved by the Saxon State Parliament. We gratefully acknowledge computation time at the Norddeutscher Verbund f\"ur Hoch- und H\"ochstleistungsrechnen (HLRN) under grant shp00026, and on the Bull Cluster at the Center for Information Services and High Performance Computing (ZIH) at Technische Universit\"at Dresden. The authors also thank Henrik Schulz and Jens Lasch for providing very helpful support on high performance computing at HZDR.

\section*{Data Availability}
{The data supporting the findings of this study are available on the Rossendorf Data Repository (RODARE)~\cite{data}.}

\bibliography{bibliography.bib}

\end{document}